\documentclass{article}
\usepackage{spconf,amsmath,graphicx, bm}

\title{High-Accuracy and Low-Latency Speech Recognition with \\ Two-Head Contextual Layer Trajectory LSTM Model}
\name{Jinyu Li, Rui Zhao, Eric Sun, Jeremy H. M. Wong, Amit Das, Zhong Meng, and Yifan Gong}
\address{Microsoft Speech and Language Group}
\begin{document}
\ninept
\maketitle
\begin{abstract}
While the community keeps promoting end-to-end models over conventional hybrid models, which usually are long short-term memory (LSTM) models trained with a cross entropy criterion followed by a sequence discriminative training criterion, we argue that such conventional hybrid models can still be significantly improved. In this paper, we detail our recent efforts to improve conventional hybrid LSTM acoustic models for high-accuracy and low-latency automatic speech recognition. To achieve high accuracy, we use a contextual layer trajectory LSTM (cltLSTM), which decouples the temporal modeling and target classification tasks, and incorporates future context frames to get more information for accurate acoustic modeling. We further improve the training strategy with sequence-level teacher-student learning. To obtain low latency, we design a two-head cltLSTM, in which one head has zero latency and the other head has a small latency, compared to an LSTM. When trained with Microsoft's 65 thousand hours of anonymized training data and evaluated with test sets with 1.8 million words, the proposed two-head cltLSTM model with the proposed training strategy yields a 28.2\% relative WER reduction over the conventional LSTM acoustic model, with a similar perceived latency. 
\end{abstract}
\section{Introduction}
\label{sec:intro}

There is a clear trend of recent development of end-to-end (E2E) modeling \cite{miao2015eesen, soltau2016neural, battenberg2017exploring, chiu2018state, he2019streaming, Karita2019Transformer, wang2019semantic, Yu-RecentProgDeepLearningAcousticModels}. However, it is still difficult for E2E models to replace the popular hybrid systems in industry, where flexibility of modeling is important. 
%For example, although Google's first large-scale E2E automatic speech recognition (ASR) system was pioneered in 2018 \cite{chiu2018state}, it is still unable to outperform their best hybrid system in a recent report \cite{Sainath19}. 
Furthermore, latency or even streaming is always a concern for E2E models \cite{Sainath19, chiu2017monotonic, moritz2019triggered}, while hybrid systems usually have low latency.  Hence, hybrid systems continue to dominate in industry, and advancing the hybrid systems is still an important research topic. Conventional hybrid systems are usually trained, first with a cross entropy (CE) criterion, followed by a sequence discriminative criterion, such as maximum mutual information (MMI) \cite{woodland2002large} and state-level minimum Bayes' risk (sMBR) \cite{gibson2006hypothesis}. Such hybrid systems \cite{pundak2016lower} were used as strong baselines to justify the accuracy advantage of E2E models \cite{chiu2018state}. However, we argue that such baseline hybrid models can still be improved significantly, retaining their competitiveness with the improvements in E2E modeling. In this study, we detail our efforts to develop high-accuracy and low-latency hybrid models. 

To achieve high accuracy, we work in two directions. The first direction is to have an advanced model structure. There have been many works to improve the dominant long short-term memory (LSTM) \cite{Hochreiter1997long, Sak2014long} model structures for acoustic modeling, such as highway LSTM \cite{HighwayBLSTM-zhang2016}, residual LSTM \cite{zhao2016multidimensional, kim2017residual}, time-frequency LSTM \cite{Li15FLSTM, Li16TFLSTM, sainath2016modeling}, and grid LSTM \cite{kalchbrenner2015grid, hsu2016prioritized}. In this paper, we use a new model called context layer trajectory LSTM \cite{Li19cltLSTM} to significantly improve the model accuracy. The advantage of cltLSTM comes from 1) decoupling the task of temporal modeling and target classification using time-LSTM and depth-LSTM respectively; and 2) exploring context frames to incorporate future information. 
%However all these models work in a layer-by-layer and step-by-step fashion. The output of a LSTM unit (either the standard time LSTM or grid LSTM) is used as the input to the LSTM  of the next layer at the same time step. The output of the final LSTM layer is used for senone classification. However, this design may not be optimal for the LSTM outputs to serve the purposes of both temporal modeling along time axis and senone classification along the depth axis. In \cite{Li18ltLSTM}, a layer trajectory LSTM (ltLSTM) was proposed to decouple the tasks of temporal modeling with time-LSTMs and senone classification with depth-LSTMs. Furthermore, the depth-LSTMs create auxiliary connections for gradient flow, thereby making it easier to train deeper models. It was shown to significantly outperform LSTM or residual LSTM. %In \cite{Li18ltLSTMExplore}, a generalized ltLSTM architecture is proposed with the concept of depth processing block which can contain any units serving for the purpose of classification. 
%In \cite{Li19cltLSTM}, the contextual layer trajectory LSTM (cltLSTM) was proposed to further improve the performance of ltLSTM models  by using context frames to capture future information. 
The second direction is to improve upon the hybrid model training strategy. Specifically, on top of sequence discriminative training, we further improve the model accuracy by performing sequence-level teacher-student (T/S) learning \cite{wong2016sequence, wong2019general} toward a strong ensemble teacher.
%The standard hybrid training recipe is to first train a cross-entropy (CE) model and then refine it with a sequence discriminative training criterion. 

To obtain low latency, we propose a two-head cltLSTM structure that has two softmax output layers with shared time-LSTM units but different depth-LSTM branches. One depth-LSTM branch does not use any future context frames, and hence has zero additional latency. This is used for first-pass decoding. The other depth-LSTM branch incorporates future context frames for high-accuracy modeling, and is used for second-pass decoding. Such a design allows for an ASR model with high accuracy and low perceived latency.

Using 65 thousand hours of Microsoft  anonymized production training data with personally identifiable information removed, our proposed model together with the proposed better training strategy achieved a 28.2\% relative word error rate (WER) reduction from the conventional MMI-trained LSTM model, while having almost the same low perceived latency. 

%This paper is organized as follows. In Section \ref{sec:prior}, We will discuss the relation to prior work. Finally, we conclude the paper in Section \ref{sec:con}.

\section{Improvements to the LSTM acoustic model}

For a multi-layer LSTM, we define the hidden output of the $l$th layer at time $t$ as
\begin{align}
{h}_t^l &= \text{LSTM}\left(h_{t-1}^l, x_t^l\right),\label{eq:temporal}
\end{align}
where the $\text{LSTM}()$ function is the standard LSTM unit with a projection layer \cite{Sak2014long}. Here, $h_{t}^l$ is the hidden output of the $l$th layer at time $t$ and ${x}_{t}^l$ is the input vector for the $l$th layer with
 \begin{equation}
    {x}_{t}^l = 
\begin{cases}
    {h}_{t}^{l-1},& \text{if } l > 1 \label{eq:lstmx} \\
    {s}_t,              & \text{if } l = 1
\end{cases},
\end{equation}
where ${s}_t$ is the speech spectrum input at time step $t$. Next, several models used in this study will be introduced

\subsection{Layer trajectory LSTM}
\label{ssec:ltLSTM}

The standard LSTM units used in recurrent neural networks serve two very different purposes at the same time, namely temporal modeling and target classification. In \cite{Li18ltLSTM}, the layer trajectory LSTM (ltLSTM) was introduced to decouple the tasks of temporal modeling and target classification, using time-LSTM and depth-LSTM units respectively. As is reported in \cite{Li18ltLSTM}, the ltLSTM significantly outperformed the LSTM or residual LSTM. The time-LSTM formulation is the same as Eq. \eqref{eq:temporal} while the depth-LSTM formulation is 
\begin{align}
{g}_t^l &= \text{LSTM}\left(h_{t}^l, {g}_t^{l-1}\right),	\label{eq:depth}
\end{align}
where ${g}_t^l$ is the output of the depth-LSTM at layer $l$ and time $t$. 

\subsection{Contextual layer trajectory LSTM}
\label{ssec:cltLSTM}
In \cite{Li19cltLSTM}, the contextual layer trajectory LSTM (cltLSTM) was proposed to further improve the performance of ltLSTM models  by using context frames to capture future information. 
In the cltLSTM,  ${g}_{t}^{l-1}$ in Eq. \eqref{eq:depth} is replaced by the look-ahead embedding vector ${\zeta}_{t}^{l-1}$ in order to incorporate future context information,
\begin{align}
%{h}_t^l &= LSTM(h_{t-1}^l,  x_t^l)	\\
{g}_t^l &= \text{LSTM}(h_{t}^l, {\zeta}_{t}^{l-1}),	\label{eq:depth-clt}
\end{align}
while ${h}_t^l$ is still computed with Eq. \eqref{eq:temporal}. The embedding vector is computed by transforming the outputs of the depth-LSTM from current and future frames as
\begin{align}
{\zeta}_{t}^{l-1} &= \sum_{\delta =0}^{\tau}  {G}_{\delta}^{l-1} {g}_{t+\delta}^{l-1}, \label{eq:cltLSTM} 
\end{align}
where ${G}_{\delta}^{l-1}$ denotes the weight matrix applied to the depth-LSTM output ${g}_{t+\delta}^{l-1}$. An $L$ layer cltLSTM with $\tau$ future context frames at each layer has a total of $N = L \times \tau$ look-ahead frames. 
%By utilizing future information, cltLSTM can achieve better accuracy than ltLSTM \cite{Li19cltLSTM}.

\subsection{Two-head contextual layer trajectory LSTM}
\label{ssec:th-cltLSTM}

\begin{figure}[t]
  \centering
  \includegraphics[width=\linewidth]{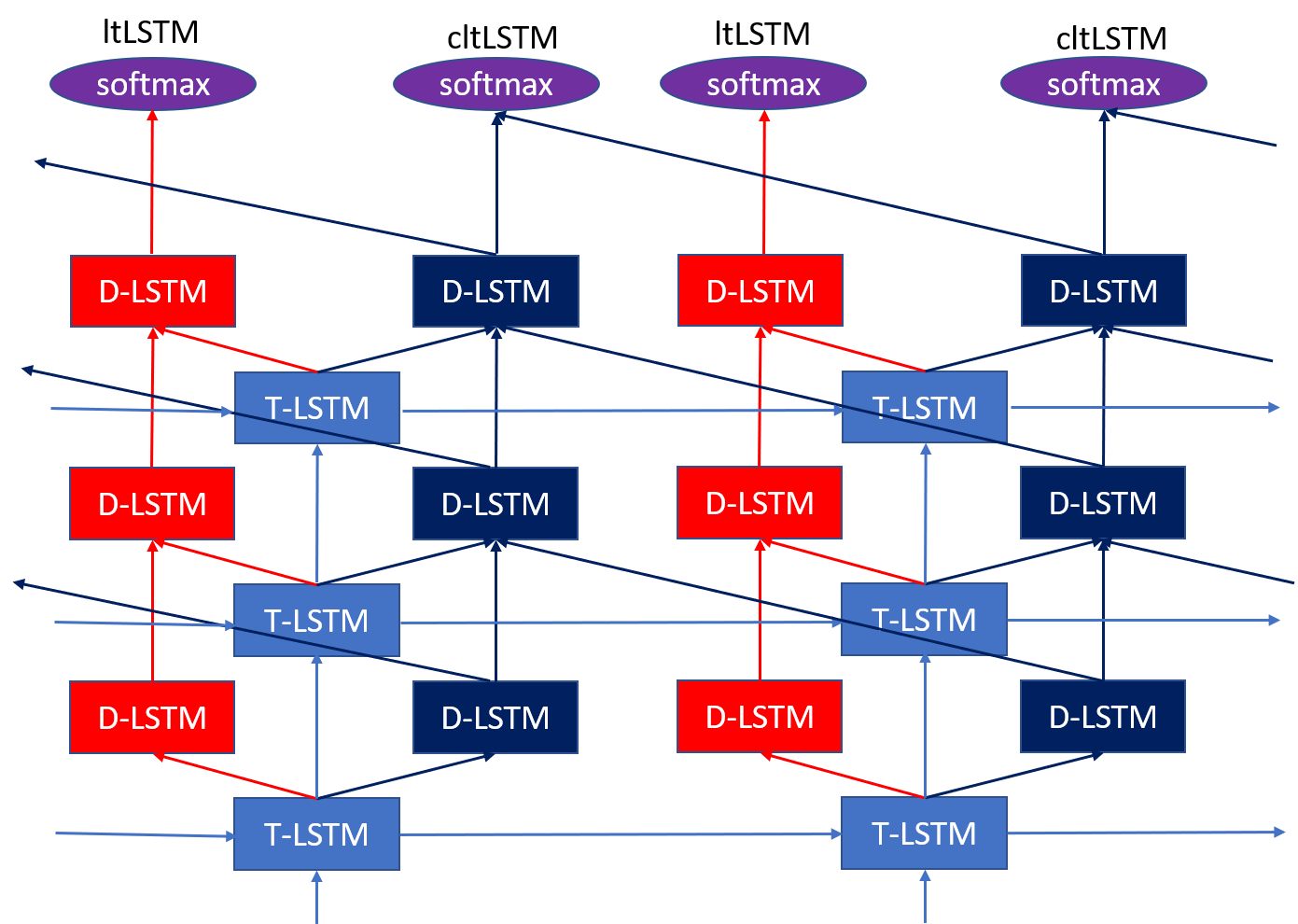}
  \caption{Diagram of the two-head cltLSTM. The time-LSTM (T-LSTM) units are shared, while the ltLSTM head and cltLSTM head have separate depth-LSTM (D-LSTM) units.}
  \label{fig:two-head}
\end{figure}

Ideally, industrial speech services should have both high accuracy and low latency. The latter is usually overlooked in many studies, but is very important to the user's experience. A high latency gives the user the impression that the system is not responding, even though it may have a high accuracy. However, these two requirements sometimes conflict with each other, especially when the system explores future information (e.g., cltLSTM) to boost its modeling accuracy. In this study, we propose a two-head cltLSTM shown in Figure \ref{fig:two-head} to build ASR systems with both high accuracy and low latency.

The two-head cltLSTM has an ltLSTM head and a cltLSTM head, which share the same time-LSTM units, but have their own respective depth-LSTM units. The ltLSTM head, without any access to future context frames, provides low-latency decoding, while the cltLSTM head provides high-accuracy recognition results, thanks to the future look-ahead. The training steps are
\begin{enumerate}
	\item Train a cltLSTM model with the best training recipe, which will be described in Section \ref{sec:ts}. 
	\item Take the time-LSTM layers out from the well-trained cltLSTM, and then build an ltLSTM with Eq. \eqref{eq:depth}, by adding the depth-LSTM and softmax output. With a new softmax layer, we form an ltLSTM head for the two-head cltLSTM. 
	\item Train the new ltLSTM model without updating the time-LSTM layers. 
\end{enumerate}

The runtime steps of the two-head cltLSTM are as follows.
\begin{enumerate}
	\item Start the first-pass decoding using the ltLSTM head. For every frame, store the time-LSTM hidden vectors ${h}_t^l$ at all layers, calculated with Eq. \eqref{eq:temporal}. 
	\item Start the second-pass decoding using the cltLSTM head after the first pass-decoding processes $N$ acoustic frames, where $N$ is the number of contextual frames that the cltLSTM model needs to look ahead. 
	\item Replace the output results of the first-pass decoding with the output results of the second-pass decoding when they differ.
\end{enumerate}
The first-pass decoder gives users almost 0 latency, while the second-pass refines the results later on. Since the absolute WER difference between cltLSTM and ltLSTM is small, the replacement of results at runtime step 3 only happens occasionally. Furthermore, a small $N$ ensures that the replacement latency is not too long. Therefore, the whole system has a high accuracy and a low perceived latency.

%\section{Sequence level T/S learning}
\section{Boosting accuracy with sequence-level teacher-student learning}
\label{sec:ts}

A conventional hybrid model training recipe is to first train toward the frame-level CE criterion, followed by a sequence discriminative criterion such as MMI \cite{woodland2002large},  sMBR \cite{gibson2006hypothesis} or the recently proposed word-level edit-based minimum Bayes' risk (EMBR) \cite{shannon2017optimizing}. 
%While our regular modeling training recipe is to train a hybrid model with the CE criterion followed by a MMI criterion, we want to explore more sequence training methods to boost model accuracy. A straight-forward way is to apply EMBR on CE-trained models, or on MMI-trained models. 
In this study, we further boost the accuracy by using teacher-student (T/S) learning to train the model to emulate a strong ensemble. 

A popular T/S learning strategy was first proposed by Li et al. in 2014 \cite{li2014learning} and later by Hinton et al. in 2015 \cite{hinton2015distilling} as knowledge distillation. The training criterion is to minimize the Kullback-Leibler (KL) divergence between the frame-level output posterior distributions of the teacher and student networks. Although it has achieved much success in deep learning, it does not take into account the sequential nature of speech. Instead, \cite{wong2016sequence} proposed to do T/S learning at the sequence level by minimizing the KL divergence between hypothesis sequence posteriors,
\begin{align}
	\mathcal{L}_\text{seq-TS}\left({\theta}_\text{S}\right) =
	-\sum_H
	P\left(H \middle|X;{\theta}_\text{T}\right) \log P\left(H \middle|X;{\theta}_\text{S}\right),
	\label{eqn:loss_sts}
\end{align}
where $X$ is an input sequence, $\theta_\text{T}$ is the teacher, $\theta_\text{S}$ is the student, and $H$ is a hypothesis, which may be expressed as a sequence of words, sub-word units, or states \cite{wong2019general}. This study considers state sequences.

Strong teacher targets may be obtained by combining an ensemble of multiple teachers together, such that
\begin{align}
		P\left(H \middle|X;{\theta}_\text{T}\right) = \sum_k \alpha_k P\left(H \middle|X;{\theta}_{\text{T}_k}\right),
		\label{eqn:hyp_combine}
\end{align}
where $\alpha_k$ is the combination weight for the $k$th teacher, ${\theta}_{\text{T}_k}$, satisfying $\sum_k \alpha_k=1$ and $\alpha_k\ge 0$. The contribution to the T/S gradient from the teachers can be obtained by performing a separate forward-backward operation over each of the teachers' lattices, which represent each teachers' hypotheses \cite{wong2019general}. This is computationally expensive, especially when using a large amount of training data. To reduce this cost, we instead combine the teachers at the frame level, 
\begin{align}
		P\left(q_t \middle|x_t;{\theta}_\text{T}\right) = \sum_k \alpha_k P\left(q_t \middle|x_t;{\theta}_{\text{T}_k}\right),
		\label{eqn:frame_combine}
\end{align}
where $q_t$ is the senone state at time $t$. The combined frame posteriors can then be used with a single lattice for all combined teachers, over which a single forward-backward instance can be performed during the gradient computation. A similar frame-level teacher combination has previously been used in \cite{kanda2017investigation} within a lattice-free framework. In this paper, we investigate its use within a lattice-based framework.
%In this way, no lattice generation from each individual network and union process in the hypotheses level is needed. 

In \cite{wong2016sequence}, a unigram word-level language model (LM) was used to generate lattices for sequence T/S learning. This followed the justification used in MMI \cite{woodland2002large}, that a weak LM should be used for lattice generation to allow for a diverse representation of hypotheses. However, in the context of sequence T/S learning, we argue that it is better to generate the training lattices using a strong LM, so that the targets are as representative as possible of hypotheses generated at runtime, to allow the student to better emulate the teacher's runtime behavior. In this paper, we will compare the use of different ngram LMs for lattice generation for sequence T/S learning within a lattice-based framework. 

The proposed hybrid model training recipe is as follows. 
\begin{enumerate}
	\item Train the  model of interest, A,  and another strong model, B, with the CE, MMI, and then EMBR criteria in order.
	\item Use a frame-level combination of the EMBR versions of A and B as the teacher ensemble with Eq. \eqref{eqn:frame_combine}.
	\item Initialise the student as the MMI version of A and perform sequence T/S learning toward the ensemble with Eq. \eqref{eqn:loss_sts}. 
\end{enumerate}

\section{Experiments}
\label{sec:exp}

In this section, we evaluate the effectiveness of the proposed models. All models were trained with 65 thousand (K) hours of transcribed data from a variety of Microsoft products. The test set covers 13 application scenarios such as Cortana and far-field speech, using a total of 1.8 million (M) words. All the training and test data are anonymized data with personally identifiable information removed. We report the WER averaged over all test scenarios. 

All of the models have 6 uni-directional LSTM layers with 1024 hidden units, and the output dimension is reduced to 512 using a linear projection layer \cite{Sak2014long}. The input feature is 80-dimension log Mel filter bank for every 10 milliseconds (ms) of speech.  The softmax layer has 9404 nodes to model the senone labels. The target senone label is delayed by 50 ms, similarly to \cite{Sak2014long}. We applied frame skipping by a factor of 2 \cite{Miao16} to reduce the runtime cost, which corresponds to 20 ms per frame. Runtime decoding is performed using a 5-gram language model with around 100 M ngrams.

\subsection{Training strategy}

We first explore how to build a high-accuracy hybrid model. We use the cltLSTM with a 24-frame look-ahead ($\tau=4$ and $L=6$) described in Section \ref{ssec:cltLSTM}, denoted here as cltLSTM-24. Since we applied frame skipping by a factor of 2 \cite{Miao16}, this 24-frame look-ahead introduces 480 ms of latency compared to a standard LSTM model. There are 63 M parameters in this model. As shown in Table \ref{tab:recipe}, the CE-trained cltLSTM-24 model has a WER of 11.15\% averaged over the whole test sets. MMI training reduces the WER to 10.36\%. EMBR training from the CE seed model yields a similar WER to the MMI model. Starting from the MMI model, additional EMBR training further reduces the WER to 10.18\%.
%additional EMBR training yields a WER of 10.18\%,  slightly improving over the MMI model. 
%which is a 1.7\% relative improvement over the MMI model. 

\begin{table}[t]
  \caption{Average WERs of cltLSTM-24 and lt-lc-BLSTM on 13 test sets with 1.8 M words.}
  \label{tab:recipe}
  \centering
  \tabcolsep=5pt
  \begin{tabular}{l|cc|c}
    	\hline
							& cltLSTM-24 &  lt-lc-BLSTM & combine   \\							
    	\hline
		CE 			& 11.15 & 10.11 & -\\
    CE $\rightarrow$ MMI 		& 10.36 & 9.45 & -\\
		CE $\rightarrow$ EMBR &	10.38 & - & -\\ 
		MMI $\rightarrow$ EMBR  & 10.18 &  9.24 & 8.62\\
		\hline
		MMI $\rightarrow$ T/S (bigram) 	& 9.63 & -  & -\\ 
		MMI $\rightarrow$ T/S (5-gram) 	& 9.34 & 8.92 & -\\ 
		
    	\hline
  \end{tabular}
\end{table}

Next, we further improve the model with sequence T/S learning toward an ensemble, as is described in Section \ref{sec:ts}. The ensemble combines the cltLSTM-24 with a layer-trajectory latency-control bi-directional LSTM (lt-lc-BLSTM) \cite{Sun19ltBLSTM}. Similarly to the layer-trajectory LSTM models used in this study, the lt-lc-BLSTM also decouples the tasks of temporal modeling and target classification with time-BLSTM and depth-LSTM units, respectively. 
It also has 6 hidden layers. At each layer, the forward and backward LSTMs in the time-BLSTMs use 800 hidden units and then are projected to 400 by a linear projection layer. The depth-LSTMs uses 800 hidden units that are also then projected to 400.  
It totally has 102 M parameters, and has up to 800 ms of latency with the latency-control implementation \cite{HighwayBLSTM-zhang2016}. The size and latency of this lt-lc-BLSTM do not satisfy the requirements for most Microsoft application scenarios. However, it is a strong model that can be combined with the cltLSTM-24 as a teacher ensemble for sequence T/S learning. From Table \ref{tab:recipe}, this lt-lc-BLSTM has WERs of 10.11\%, 9.45\%, and 9.24\% for CE, MMI, and EMBR training, respectively. Using Eq. \eqref{eqn:frame_combine}, an equally weighted frame-level combination of the EMBR cltLSTM-24 and lt-lc-BLSTM models yields a teacher ensemble WER of 8.62\%. 

The cltLSTM-24 MMI model was used as the initial student. We experimented on generating the lattices that represent the T/S hypotheses using either a bigram or 5-gram LM, with the acoustic scores from the initial cltLSTM-24 MMI model. The lattices were acoustically re-scored with the frame-level combined teachers' acoustic scores to compute the contribution to the T/S gradient from the teachers. The 5-gram LM was the same as that used at runtime. The cltLSTM-24 student trained using the bigram and 5-gram lattices achieved WERs of 9.63\% and 9.34\%, respectively. This shows that sequence T/S learning performs better when the hypotheses are represented with the stronger LM that is used during runtime. The final cltLSTM-24 student yields an 9.8\% relative WER improvement over its MMI initialization. As a comparison, an lt-lc-BLSTM student, %starting from the MMI initialisation, 
was also trained toward the same ensemble, yielding a WER of 8.92\%. This is better than the cltLSTM-24 student, but has a model size and latency that is too expensive for runtime application. %It is shown in \cite{wong2019different} that it may be difficult for a student to effectively emulate the teacher's behaviour when it has a shorter temporal context, with experiments conducted using training sets with around 100 hours of data. The results here suggest that this may also be true when using 65K hours of training data.
%Using lattice generation with 5gram and the same teacher model, he lt-lc-BLSTM student model improved from its MMI counterpart with 5.6\% relative WER reduction.

\subsection{Models with different look-ahead frames}
The next experiment used the same CE$\rightarrow$MMI$\rightarrow$T/S training strategy, using the same teacher ensemble as Table \ref{tab:recipe}. The students used were LSTM, ltLSTM (zero frame look-ahead), and cltLSTM-6 and cltSLTM-12, which have 6 ($\tau=1$) and 12 ($\tau=2$) frames look-aheads. The WERs of these models are reported in Table \ref{tab:wer}. The latency and the number of parameters are shown in Table \ref{tab:runtime}. For all models, consistent gains are observed from CE to MMI, and then to sequence T/S. The relative WER reductions from sequence T/S over MMI for LSTM, ltLSTM, cltLSTM-6, cltLSTM-12, and cltLSTM-24 are 11.7\%, 8.7\%, 9.5\%, 9.5\%, and 9.8\%, respectively.  %Hence, the two-sage  (MMI and T/S) sequence training is the key to generate models with high accuracy. 

Comparing the final T/S models, the 12.1\% relative WER reduction of ltLSTM over LSTM shows the benefit of decoupling the temporal modeling and target classification tasks. The 4.4\% relative WER reduction of cltLSTM-6 over ltLSTM indicates the effectiveness of incorporating future information. cltLSTM-12 further reduces the WER by 3.3\% relative over cltLSTM-6, while cltLSTM-24 does not yield further gains. The frame skipping \cite{Miao16} used during runtime means that every frame spans 20 ms. Therefore, ltLSTM, cltLSTM-6, cltLSTM-12, and cltLSTM-24 respectively have 0, 120, 240, and 480 ms greater latencies than an LSTM. It is therefore better to use cltLSTM-12 rather than cltLSTM-24, since both yield similar WERs but cltLSTM-12 has half the latency. However, the 240 ms latency of cltLSTM-12 may still result in poor user experience. The next section considers the two-head cltLSTM model to alleviate this. 

\begin{table}[t]
  \caption{Average WERs of all models on 13 test sets with 1.8 M words.}
  \label{tab:wer}
  \centering
  \begin{tabular}{l|c|c|c}
    	\hline
				models			& 	CE & 			MMI & sequence  T/S        \\							
    	\hline
		LSTM 			& 14.75 & 13.01 & 11.49 \\
    ltLSTM      & 12.41 & 11.06  & 10.10 \\
		cltLSTM-6 &	11.97 & 10.67 & 9.66 \\ 
		
		cltLSTM-12  & 11.38 & 10.32 & 9.34 \\
		cltLSTM-24 	& 11.15 & 10.36 & 9.34 	 \\ \hline
		
		two-head cltLSTM-12 & & & 	\\
		first head   	& 12.24 & 11.33 & 10.03	 \\
    second head   & 11.38 & 10.32 & 9.34	\\ 	 
    	\hline
  \end{tabular}
\end{table}
\subsection{Two-head cltLSTM}
We extracted the time-LSTM out from cltLSTM-12, and built ltLSTM on it, as is described in Section \ref{ssec:th-cltLSTM}. We then do CE, MMI, and sequence T/S training for this new ltLSTM without updating the time-LSTM parameters. The WERs and costs of this two-head cltLSTM are shown at the bottom of Tables \ref{tab:wer} and \ref{tab:runtime} respectively. The first head, ltLSTM with zero additional latency compared to LSTM, obtained a WER of 10.03\% WER after sequence T/S learning, which is slightly better than the 10.10\% obtained by the ltLSTM model trained from scratch. The second head, cltLSTM-12, is the same model as the cltLSTM-12 trained from scratch. In the first-pass decoding, we run a decoder with the first head. After 240 ms, we kick off the second-pass decoding. Because the WER gap between ltLSTM and cltLSTM-12 is only 0.69\% absolute, just a small fraction of the words are replaced from the first-pass results by the second-pass. %Furthermore, since the second-pass decoding only has a 240 ms delay from the first-pass
Therefore, the perceived latency is small. 

\begin{table}[t]
  \caption{Latency and number of parameters of all models.}
  \label{tab:runtime}
  \centering
  \begin{tabular}{l|c|c}
    	\hline
							& 	 latency compared & number of\\ 
							& to LSTM (ms) &  parameters (M)    \\							
    	\hline
		LSTM 			& 0 & 31 \\
    ltLSTM      & 0 & 57 \\
		cltLSTM-6 &	120 & 58 \\ 
		
		cltLSTM-12  & 240 & 60 \\
		cltLSTM-24 	& 480 & 63\\ \hline
		
		two-head cltLSTM-12 & & 	\\
		first head   	& 0 & 57 \\
    second head   & 240 & 34 	\\ 	 
    	\hline
  \end{tabular}
\end{table}

Since the time-LSTM units are shared in the two-head cltLSTM-12, the total number of parameters is $57+34=91$ M. We can also use an LSTM in the first-pass decoding and a separate cltLSTM-12 in the second-pass decoding. This setup also has $31+60=91$ M parameters. However, the LSTM only has a 11.49\% WER, much worse than the 10.10\% WER obtained from the first head ltLSTM in the two-head cltLSTM-12. When compared to a conventional baseline hybrid setup \cite{pundak2016lower} that trains an LSTM with the CE and then the MMI criteria, the two-head cltLSTM-12 reduces the WER from 13.01 to 9.34\%, which is a 28.2\% relative reduction. While we are also working on replacing hybrid models with E2E models \cite{rnnt-ms}, the work conducted in this paper indeed presents us a super challenging hybrid model baseline to beat. 

\section{RELATION TO PRIOR WORK}
\label{sec:prior}

Like the cltLSTM, the grid LSTM \cite{kalchbrenner2015grid, hsu2016prioritized} also operates along both the time and depth axes. However, the grid LSTM works in a layer-by-layer and step-by-step fashion, while the cltLSTM totally decouples the temporal modeling and target classification with dedicated time-LSTM and depth-LSTM units. Furthermore, the cltLSTM has the context modeling which leverages more information from future context frames, while the grid LSTM does not. Unlike the grid LSTM which cannot be configured to handle heads with different latency requirement, the decoupled temporal modeling and target classification allows for the two-head cltLSTM model design, which is shown to have low perceived latency and high accuracy.

We improve upon sequence T/S learning \cite{wong2016sequence} by using a frame-level teacher combination and a strong LM for lattice generation. Using frame-level combination of the teachers and a strong LM have already been investigated for lattice-free sequence T/S \cite{kanda2017investigation, manohar2018teacher}. In this paper, these are studied in a lattice-based implementation, and also with a large training set. We also innovated the training process as  CE$\rightarrow$MMI$\rightarrow$T/S  to get the best model accuracy.

Previously investigated two-pass decoding usually starts the second-pass after the first-pass decoding finishes \cite{Sainath19}. In contrast, the proposed two-head cltLSTM model starts the second-pass decoding just 240 ms after the first-pass decoding starts in the cltLSTM-12 setup. This yields a low perceived latency and high final accuracy. 
%The two-head cltLSTM model enables us to deliver a system with very small perceived latency. Also different from the most recent two-pass end-to-end work \cite{Sainath19} which performs second-pass rescoring after first-pass results are generated, we directly do the second-pass decoding with only 240 ms delay after the first-pass decoding starts.

\section{Conclusions}
This study aims to achieve high-accuracy and low-latency ASR by designing a two-head cltLSTM model, with one ltLSTM head for first-pass decoding and another cltLSTM head for a second-pass. The ltLSTM has the same latency as an LSTM, but has improved accuracy, due to decoupling of temporal modeling and senone classification tasks. The cltLSTM further improves upon the ltLSTM by using context frames to incorporate future information. This design enables high-accuracy and low perceived latency performance. Improvements to lattice-based sequence T/S learning were also investigated, by simplifying the teacher combination and using a strong LM, to allow the student to better emulate the teachers' runtime behaviour. When trained with Microsoft's 65 K hours anonymized training data, the proposed two-head cltLSTM model and new training strategy yield a 28.2\% relative WER reduction from an LSTM trained with the conventional CE then MMI strategy, while retaining a perceived latency that is similar to the LSTM. 
\label{sec:con}
\bibliographystyle{IEEEbib}
\bibliography{refs}

\end{document}